# TALON - The Telescope Alert Operation Network System: Intelligent Linking of Distributed Autonomous Robotic Telescopes

R.R. White*, J. Wren, H. Davis, M. Galassi, D. Starr, W.T. Vestrand and
P. Wozniak
Los Alamos National Laboratory, Los Alamos, NM 87545, USA

## ABSTRACT

The internet has brought about great change in the astronomical community, but this interconnectivity is just starting to be exploited for use in instrumentation. Utilizing the internet for communicating between distributed astronomical systems is still in its infancy, but it already shows great potential. Here we present an example of a distributed network of telescopes that performs more efficiently in synchronous operation than as individual instruments. RAPid Telescopes for Optical Response (RAPTOR) is a system of telescopes at LANL that has intelligent intercommunication, combined with wide-field optics, temporal monitoring software, and deep-field follow-up capability all working in closed-loop real-time operation. The Telescope ALert Operations Network (TALON) is a network server that allows intercommunication of alert triggers from external and internal resources and controls the distribution of these to each of the telescopes on the network. TALON is designed to grow, allowing any number of telescopes to be linked together and communicate. Coupled with an intelligent alert client at each telescope, it can analyze and respond to each distributed TALON alert based on the telescopes needs and schedule.

## 1. INTRODUCTION

A series of telescopic systems can be combined using the same engineering as a distributed sensor network (DSN). In some cases the process is easier due to the use of hard wired communications and the network size tending to be small. Currently telescopes usually work as single autonomous systems where the data is analyzed and utilized on site. Combing data with other systems occurs much later, often by hand. However, it is possible by using some of the new robotic telescopic systems to analyze the data on site and transmit it to a central distribution center where it is combined and classified, then redistributed for follow-up observations.

Intelligent autonomous robotic observatories can not only take an image but can process it, calibrate it and control the observatory as a whole. These intelligent sensors could function like a colony of individual ants that, when formed into a network, cooperatively accomplish complex tasks and provide capabilities greater than the sum of the individual parts.[1] There are several advantages to linking together a group of these intelligent telescopes. By providing redundant observational areas there is less chance of losing information on a particular area (fault tolerance). A much greater observational area can be covered by combining viewing regions (mosaic coverage). Resources can target the same area to provide a greater coverage of a particular object (depth of data). By monitoring areas over successive periods of time they can provide a better characterization of continuous phenomena (temporal coverage).

Unidirectional network systems using internet communications have, for the past decade, supported satellite detector systems. Substantial effort has gone into networks that can trigger ground based optical resources to provide follow-up observations. These proved very valuable in the first identification of a gamma-ray burst's (GRB) optical counterpart, GRB 990123, imaged by the Robotic Optical Transient Search Experiment (ROTSE) telescope in response to a trigger from the BATSE satellite. But this system works in one direction. If the ROTSE system had been able to then locate the transient in its own image and trigger other systems, sending them the object coordinates, then more resources could have been brought to bear in the first few seconds of the burst. These gamma-ray bursts are important astronomically, but may not be triggering satellites due to off-axis beaming although there might be optical flashes. This shows the importance of being able to provide all-sky monitoring[2] for this and other similar types of phenomena. This has created a push during the past decade toward the creation of wide field telescopic systems. This wide-field of view combined with the new sophisticated software algorithms allows many of these new generation telescopes to monitor and note changes in the sky. In a similar way to the satellite-telescope response, these systems, when interconnected could provide fast follow-ups and independent verification of phenomena. If interconnected as an intelligent distributed sensor network the telescopes could actually communicate the desired types of instrumentations needed for follow-up and verification and the other systems could accept, reject, or re-direct the request without human observers needing to support the operation.

---

*rwhite@lanl.gov; phone 505-665-3025; fax 505-665-4414; raptor.lanl.gov

In the following sections we will discuss the intelligent distributed network interconnection method of the Telescope ALert Operations Network (TALON) and then look at the specific case of its integration into the RAPid Telescopes for Optical Response (RAPTOR) system. While the distributed telescopes of RAPTOR established the original guidelines for the network's capabilities, it has evolved into a general architecture that is useful, if not essential, to other telescope systems.

## 2. TALON - THE TELESCOPE ALERT OPERATION NETWORK

When initial planning for TALON was underway, it was noticed that the communication system could be constructed in the same way distributed sensor networks are formed. The ultimate goal of any distributed sensor network is to make decisions or gain knowledge based on information fused from distributed sensor inputs[3]. Each telescope system acts as a different input, performing calibration and image analysis on site. The reduced image information is then passed to a central system where it is combined with data from other telescopes and further analysis is performed. Software algorithms then make decisions concerning follow-up observations based on the fused data. The system then transmits follow-up alerts to all telescopes in the system, whereby they respond and begin observations of any targets of interest.

Planning dictated that if the system was to operate autonomously then the code needed to be optimized, robust and well debugged prior to operation. Connection protocols in the system had to be self diagnosing and needed the capability of searching for and reestablishing contact when needed. The code needed to be modular and portable across different operating systems and computer hardware, providing flexibility and removing any restrictions future hardware changes could impose. Because of the simplicity of the TCP/IP protocol and overall stability of Unix/Linux, this was chosen as the default operating system for the central server. All TALON programs are built using C and C++ modules following an object oriented design principle. The program is built so code modules can be removed or added as needed to support expanding or refining the code for different requirements.

The TALON code consists of three main programs. One of these is the alert client (TALON client) that runs on each telescope control computer. This provides the communication to and from the central server (TALON central) located on a remote computer. A key element for watching the operations of the network is facilitated by the TALON monitor program. This monitor program allows any user subscribing to the network the capability of seeing the current status of all telescopes connected to TALON. The system information is displayed through an easy to read GUI allowing the user to not only identify connection status but in some cases, the current observational program of the client telescopes.

**2.1 TALON Central**
A central challenge to building any advanced sensor network will be the development of robust control for networks that can scale in size.[4] Therefore the heart of the TALON system is the network protocol code utilizing scalable TCP/IP network connections. When TALON central starts it spawns four separate server systems. One of these is dedicated to incoming client communications on dedicated sockets, the second is for transmitting alerts through a single socket, the third is for outside network connections, and the fourth is for outside monitoring of the telescope/server system. Each has been broken into shared C++ classes devoted to socket generation, client membership, and client monitoring. For client membership and monitoring, the base class contains the alert transmission and client information. Inheriting from this base is the incoming data class. This class provides the methods for receiving data from passive clients (site information only) or active clients (image data and site information). The client membership code allows any number of clients to be connected through a single socket, providing full scalability for the system. This allows for multiple imaging systems from a single array to be piped down the same socket, or any number of clients to listen for alerts. A networking class cycles through all listening clients to be sure each receives the alert information. The incoming data clients are assigned dedicated sockets for transmitting data to TALON central. The dedicated socket method was chosen because it was the simplest and easiest method of determining the data origin. This also allows each client's incoming socket and monitor to be spawned off as a different thread, making the process of handling massive amounts of data from many sources to be run in parallel. A final network class is used to supply data to the TALON monitoring software. This is similar to the alert transmission module but has special control and monitoring code to pass on site information received by TALON central.

At this time TALON is being fully utilized by the RAPTOR system. In the current RAPTOR scheme, only A and B transmit data to the system for stereoscopic comparison, while all RAPTOR telescopes are connected to the alert transmission socket. The GCN network is currently the only outside network passing information into the system, but the incoming network server is coded to allow additional connections; also fully scalable. The RAPTOR telescopes are configured to return operational information back to TALON central for monitoring.

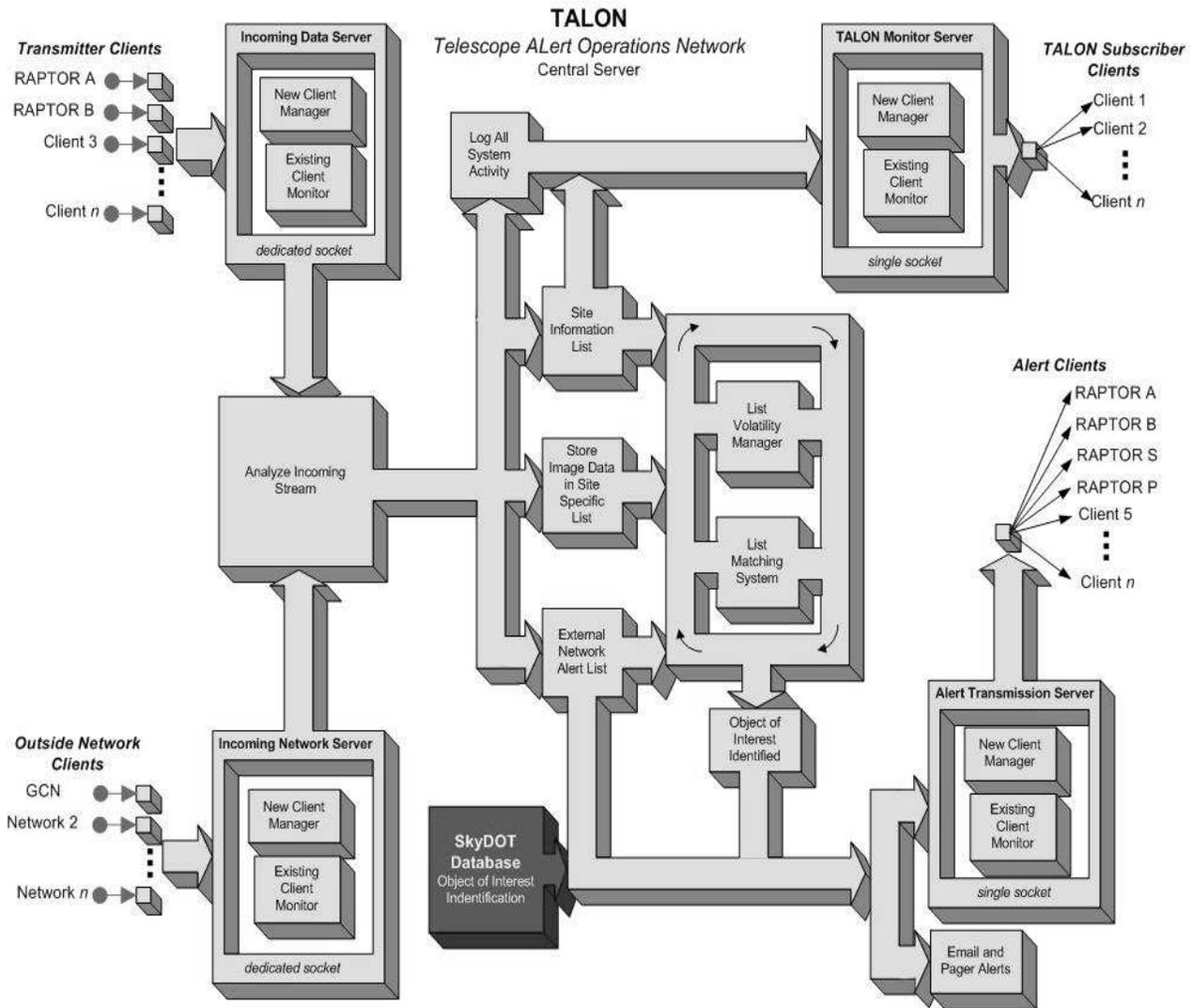

Figure 2-1 Block diagram of operational flow for TALON central.

### 2.1.1 The Flow of Information
Data arrives at the TALON central software through robust, fault tolerant socket connections. Once a socket is created its activity is monitored via the standard `select()` function. This function, through careful coding and error checking, reveals connection, disconnection, read activity, write activity and the information source. Connection and disconnection situations are handled by the client membership functions; sending or receiving "I'm alive" packets to (or from) the client is not necessary. Incoming data packets (through the dedicated sockets), are immediately decoded and the message header identifies the type of information. Currently the RAPTOR telescopes are only issuing two types of information packets; image object lists and site information. Site information is not stored, but is passed directly to the outgoing socket for the TALON monitor. Image object list packets are sent to decoding modules dedicated to each of the telescopes. Incoming networks such as GCN are also given their own decoding modules and as other networks come on line, these can be connected easily into TALON.

After reception, the data is decoded and stored for later analysis. Object data received from the active transmitting clients is broken down into its component parts: RA DEC, magnitude, magnitude error, time imaged and image file name. To this is added a system "time received" stamp. All the information is then placed in a linked list of structure elements containing candidate objects, one list per source socket. Thus RAPTOR A has its own candidate list and RAPTOR B has its own. The listing module is easily expandable to any future systems. This listing system is the key to the stereoscopic observation/instrument coincidence element of RAPTOR (discussed later). When new data is stored in a candidate list a comparison is made between objects in any opposing list. If there are no matches the item remains in the list for later comparisons, however if there is a match then the object information is passed to the alert transmission list. Matches are dictated by comparing RA, DEC, and imaging time. All these parameters must be within a margin of error. Currently we are using margins of 54 arc seconds in RA and DEC (approximately two pixels in the wide field optics) and 300 seconds in imaging time. These parameters may seem very loose, but the system still only generates less than two (and usually zero) follow-ups per night. Each linked list is volatile; elements (single object component parts and time stamp) can expire. When the age (stamped system time) of an object exceeds three times the imaging time error margin (900 seconds), the element is removed from the candidate list. An internal routine checks through all the available lists every sixty seconds, looking for expired elements. During the testing phase of development, the server was inundated with 10000 candidates per camera every thirty seconds (five cameras per RAPTOR transmitter) and the list modules were able to perform all comparisons within five seconds. At its maximum point the system was storing and comparing $3 \times 10^6$ objects. In practice we only generate a few tens of objects per image, so the system has never been pushed to that limit while in operation.

If a candidate is matched then the data is passed into the alert management module. It is first checked against any other objects that might be in the alert list. If there are any other items matching this candidate in the alert list, then it is not stored for transmission. This keeps down the possibility of redundant alerts being sent out to the network. Even alerts from incoming networks have to pass a comparison with the alert list, and are stored in the list or rejected. There always exists the possibility that RAPTOR might already be imaging the object when a satellite trigger is issued for that same position, or the reverse situation could happen. Incoming network alerts can have a variety of alert priorities based on position confidence and higher priority alerts will be passed on to the listening clients rather than being deleted as a redundant alert. Once a candidate is stored in the alert list, the object data is packaged and transmitted to all listening telescopes clients, transmitted as a pager alert, transmitted as an e-mail alert, and logged to a server operations text file. The object will remain in the alert list for one hour. This time parameter can be changed as part of the server configuration file. After an hour the alert is purged from the alert list. In the future, a connection will be made from the alert transmission module to the Sky Database for Objects in Time-Domain (SkyDOT), which will provide targets of opportunity alerts, based on analysis of temporal characteristics (flaring, occultation, etc.) of objects. For further information on SkyDOT, see skydot.lanl.gov .

**2.2 TALON Alert Client**
Each telescope site utilizes a TALON client to receive data from TALON central and to analyze and filter the incoming alerts. It can run as a simple operational process or it can be a daemonized process launched separately or as part of the overall operational software for the client observatory. The software is open source and can be adjusted to fit the necessities of any client as long as the transmit and receive protocols are left intact. The TALON client can be obtained by inquiring through the RAPTOR website (www.raptor.lanl.gov) or the email contact on page one of this text. It is necessary to configure the server to accept the client connections; therefore the TALON client is not available for open download through an ftp or web site. Once we are contacted, we can set up the server to accept the desired connections and we will then make the TALON client available for download.

Each The TALON client's first task is to facilitate the connection back to TALON central, but the key to the client program is a shared resource file that stores the incoming data and information. This file is referred to as the central repository. Variables and structures for containing site information and incoming alert data are placed here and accessed by the existing observatory / telescope software and the TALON client. A simple text parsing program is all that is needed to feed the alert information back into the client telescope's existing software. A future version of this program uses a socket method for transmitting and receiving data from the existing observatory / telescope software. In this version the system still provides storage, monitoring and filtering of the alert information and passing of data to the server. The user would still need to make the necessary connections to the client so that information can be passed. But all connections are local to the site, with the TALON client doing all of the connections to TALON central.

There may be some cases where perspective users may not wish to utilize the TALON client or they have existing transmit and receive programs in place. In either case we can provide the perspective user with the transmit and receive protocols necessary for them to establish their own connections to and from TALON central.

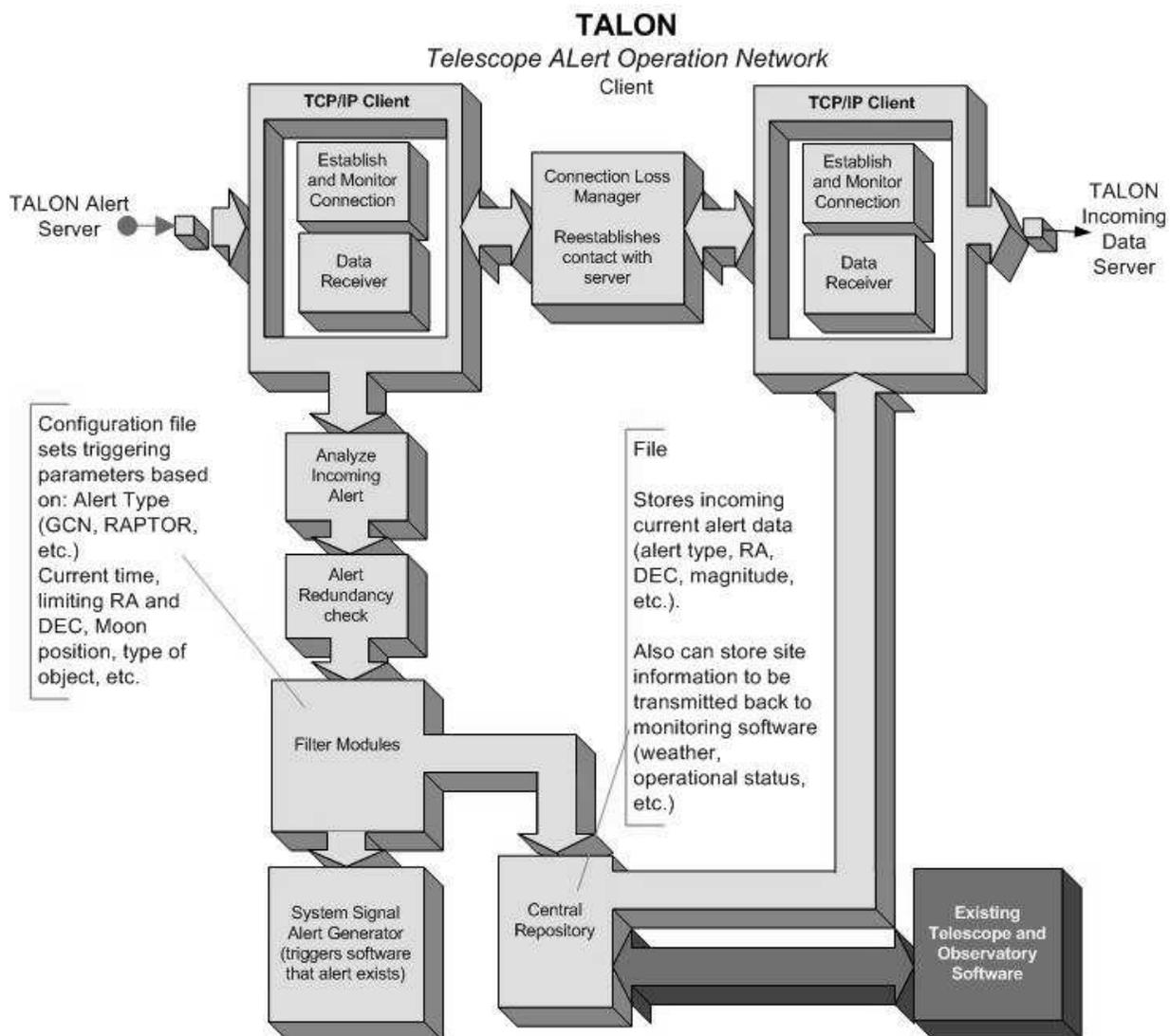

Figure 2-3  Block diagram of the operations of the TALON client.

The TALON client initially spawns two processes, depending on configuration, to control connection and monitoring to TALON central. These processes have a back up function that monitors the connection with the TALON central host.

Should a connection be lost, then that process will search and attempt re-connects at configuration defined intervals. If the transmit side was lost then data to be transmitted can be backed up in a volatile queue, to transmit when the connection is re-established. The queue makes sure that only the most recent data is sent in case the reconnection takes a long time. If the listening side is lost, then any data that might have been sent during the interval will not be transmitted when re-connected. The TALON client can be configured to only receive data (passive client), then no site information or data is sent back to the server.

**2.2.1 Following the data**
When an alert comes into the system it is filtered, decoded into its components, and placed in a temporary structure for analysis. The first step is to filter the incoming alert to see if the system is currently following up on an alert of higher priority. The priority parameters are set in the configuration file and are associated with alert types as defined by the user. In the case of the RAPTOR system, the GCN alerts are given priority and these are further filtered in a module specifically designed for GCN. Next, the client will check any alerts in operation and compare the current alert's RA and DEC with the new one. If the new alert, regardless of higher priority, would place the object's position within the field of view of the client system's telescope, then the alert will be rejected due to the fact the new alert's target is already being imaged. Additional filters for the incoming alerts can be added so that the system will consider additional variables (moon position and phase, cloud conditions, etc.) if the current observatory software does not. Currently there are only two main types of alerts that pass through the RAPTOR system. The first is the GCN based satellite alerts and the second is RAPTOR transient object follow-up alerts. In the future, target of opportunity alerts from SkyDOT will be added to the alert list. When additional telescope clients join TALON and identify and object of interest particular to their observations (i.e. supernovae, cataclysmic variables, planetary searches, etc.), their system can issue an alert, seeking support from other TALON clients. If another client telescope is particularly interested in this class of objects or has instrumentation of great use for supporting the observations, then this client will respond and notify the triggering telescope that it is also performing support observations. If an alert passes all the filters then the information is stored and the TALON client triggers the observatory software that an alert has been received. All incoming alerts are logged for later review.

When TALON clients are configured to return either data and/or site information, the repository has an alert list and site information structures for returning that data. The TALON client monitors a flag indicating that object information is available; users will need to set this flag in their code to trigger the transmission of data. When the flag is set then the TALON client will transmit the full list to TALON central, deleting objects from the list as they are transmitted. Site information is transmitted every minute to TALON central unless an object list transmission is underway. In the future version where socket transmission from the existing observatory software is used, then data is transmitted back to TALON central as it is received by the TALON client

**2.3 TALON Diagnostic system**
Being able to monitor the performance of the entire operation is a valuable resource to any multi sensor system. Each client can report back to the system any of the alarms affecting the observatory's ability to perform observations. However, the monitor is completely passive software; it has no functionality other than to display current states of the clients. No commands can be issued back to any client from the monitor. GUI design principles were used to keep the display concise and easy to read. It is fully cross platform with code functional in Linux and Windows operating systems with conversion to Mac also possible. The GUI is developed using Trolltech's QT tool set. QT is available as freeware on the Linux/Unix platform by going to www.trolltech.com. Licenses must be purchased for QT development on Windows and Mac platforms.

The diagnostic monitor gives instantaneous feedback concerning the activity and current state of client telescopes and observatories. For RAPTOR, each observatory has its own independent weather station, dome control and camera control system, and each of these can generate alarms that can halt subsystem operation. The state of each alarm is displayed by an LED type display object, with a red or green color designating the current state. Each RAPTOR also reports its state of connection to the TALON alert server, whether transmitting or listening. Other fields indicate camera imaging status, current scheduled activity (patrol, stand-by, target of opportunity, etc.) and the current pointing direction (RA, DEC) of the telescope. Some clients may not wish to follow the RAPTOR paradigm and broadcast the observing data or alarm information to TALON central. In that case only the connection status of the observatory will be seen. The window also receives constant information being stored in the TALON central log files. On the left side of the window is a text showing the connection state of the outside (incoming) networks and the last received alert from that network. The central window displays the information being logged at the server. Some of the types of data that can be seen in the server log are: connection and disconnection of clients along with time; candidates being transmitted to TALON central; match information for follow-up alerts; alert class information; and a copy of incoming network alerts. A final window shows any current errors being generated by the TALON central software or its host computer.

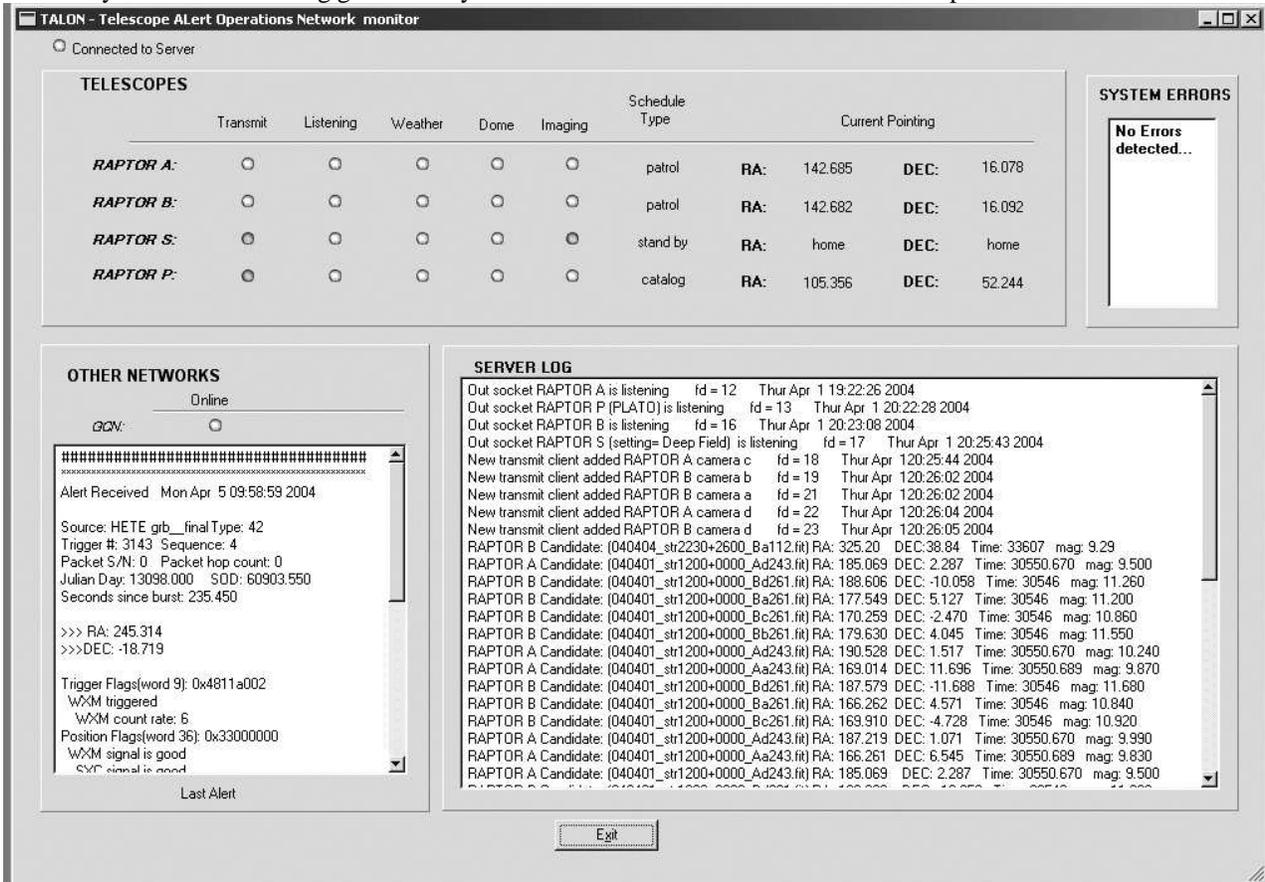

Figure 2-4  TALON monitor system GUI

## 3. USING TALON TO MERGE THE RAPTOR SYSTEM

TALON was conceived in order to join together the individual parts of RAPTOR so they could operate as a single unit, but building robotic telescopes is not an easy process. Creating a single autonomous robotic telescope is a daunting task, creating a system of them is intimidating. Not only must it be capable of scheduling its own observations, but it must robustly handle things such as changing weather conditions, software errors and hardware failures[5]. RAPTOR grew out of the lessons learned in the development of the ROTSE system, also originally located at LANL. ROTSE is a system designed to operate as a triggered telescope, but also performs patrol observations. The images from the sky patrols were analyzed after the fact, as the situation dictated. This also provided a valuable database of temporal information for the night sky over a year. RAPTOR takes the ROTSE design several steps further. RAPTOR is designed to filter out the myriad of non-celestial signals that can mimic celestial transients in order to identify the real targets[6], perform this analysis in real-time, and initiate follow-up observations of targets. TALON would prove to be an integral part in meeting the design requirements of RAPTOR.

### 3.1 Overview of the RAPTOR hardware system
RAPTOR is a system of autonomous robotic telescopes, some dedicated to monitoring and patrol and some to follow-up observations. The two main systems are RAPTOR A and B, separated by a 38km baseline. Each of these telescopes consists of an array of wide-field ($36^0 \times 36^0$ area) optic components and a narrow field ($4^0 \times 4^0$ area) fovea. Each of the five cameras on each telescope is controlled by a dedicated computer system. Each computer is responsible for controlling and monitoring the operation of the camera and to perform the pipeline analysis of the acquired image. The computer then communicates any candidate targets to the TALON. RAPTOR also contains two additional telescopic systems. RAPTOR S is a large (16") deep field instrument capable of spectroscopic or photon counting studies, operating as a follow-up instrument for alerts. RAPTOR P (PLATO – Planetary Telescope Operation) uses an array of fovea optics, similar to the fovea of RAPTOR A and B, and is currently performing solo sky patrols and responding to alerts.

The components of the RAPTOR system provide a flexible system for all-sky monitoring and fast follow-up observations with a strong depth of data for a variety of astronomical events. RAPTOR A and B provide sky patrol images watching for temporal variations in the night sky, specifically optical transients. Raptor S can provide deep field follow up and spectral or photon counting information during the earlier seconds of astronomical phenomena such as gamma-ray bursts. This information can provide valuable data about the early event processes. RAPTOR P's focus is on extra-solar planetary searches. Although extremely difficult, it can be done through analysis of light curve data as long as the system is sensitive enough and the RAPTOR P optics and system has been optimized for this purpose. In alert response mode RAPTOR P will provide additional coincidence observations matched with those of the fovea of RAPTOR A and B.

### 3.2 The real-time analysis pipeline
The real-time analysis pipeline uses a diverse collection of components and algorithms, ranging from data acquisition to source extraction, astrometry, relative photometry corrections and the transient detection algorithm[7]. The RAPTOR pipeline takes the raw image, corrects it, correlates it and then filters out a large number of the false positives from the image, producing a table of objects identified by RA, DEC magnitude, magnitude error, and time imaged. The system is quite fast, reducing the images in less than 10 seconds[7]. The pipeline runs on the RAPTOR A and B camera control computers, providing a dedicated, separate image analysis for each camera. The system is also very efficient, with an average throughput of $10^5$ objects per image; it can reduce the number of possible candidates or false positives to less than a hundred.

A large number of false positives would be an enormous drain on observing resources, so additional solutions had to be found. Follow up observations are programmed to run for an approximately ten minutes in order to obtain a statistically significant number of points on the light curve for the target object. But if each image produces a few tens of false positives, it would mean hours of useless follow-up observations. In order to reduce the possibility of false positives even further, instrumental coincidence was an excellent solution. This decision led to the creation of the dual wide-field systems. It was realized that using two systems (or more) separated by several kilometers could provide synchronized verification of target objects and parallax that could be used to eliminate nearby objects (i.e. satellites). Parallax is instrument and separation limited, so the RAPTOR wide-field cameras give a visible parallax range of approximately $3.8 \times 10^5$ km (average Earth-Moon distance). Coincidence and parallax reduces the number of false positives down to less than two a night, with most nights having no false positives.

### 3.3 Tying it all together
TALON provides the connection between all of the RAPTOR telescopes and communication with the outside world. The output of each of the camera computers in RAPTOR A and B sends a possible candidate list to TALON central where the information is stored, sorted and analyzed. This follows the DSN ideals for collaborative signal processing performed after each sensor has performed initial processing[3]. In the TALON central program, comparisons between the candidate lists are made and if any matches occur they are relayed back to the RAPTOR system as an alert. All RAPTOR telescopes listening to the network will then respond and begin making follow-up observations.

The data stream from each of the RAPTOR systems to TALON central is a compressed version of information from each image. These simple object information files are transmitted as a stream of small packets consisting of six long variables and a character string. The first long variable is the packet header, the next five are composed of the object information, and the character string contains the image fits file information.

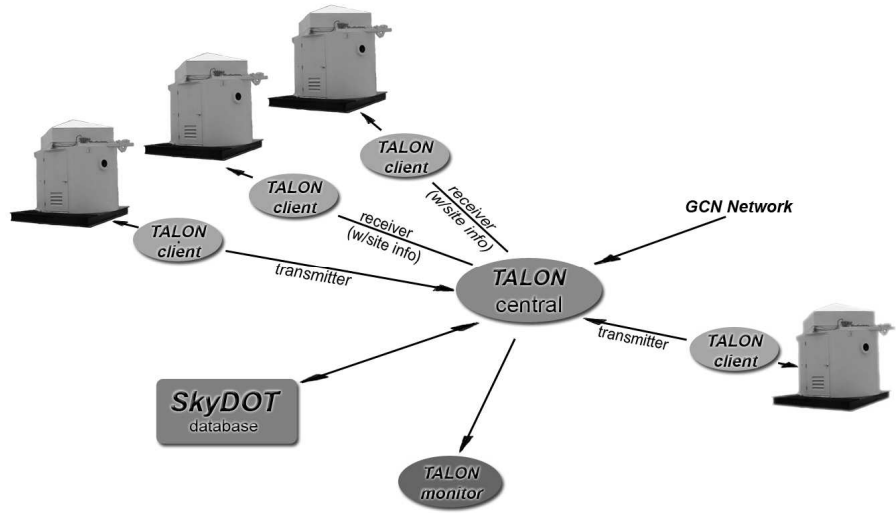

Figure 3-1: Diagram of the communications layout of the RAPTOR system. Each telescope camera computer has a TALON client running in transmit mode only. Each observatory has a TALON client running on it in listen mode for alert response. Note the bi-directional connection to the Sky Database for Objects in Time-Domain (SkyDOT), providing an evolving database of temporal information on the night sky.

### 3.4 Validating the closed-loop concept

While RAPTOR has yet to catch the optical transients it is searching for, it has performed admirably in responding to satellite triggers. The TALON loop response time, from receipt of an outside alerts until the system is on target, averages five seconds but in some cases it has been as low as three. GRB 021211 was the first GRB observed by RAPTOR, capturing the earliest optical images of this burst. RAPTOR successfully imaged the GRB in a 60 second exposure during the first two minutes of outburst[8]. Operational proof of the closed loop scheme came with the triggering of the system by asteroid Hispania 308. The RAPTOR pipeline is optimized to find new objects in the field of view. The asteroid was moving slowly enough not to show up as a linear structure on the wide field system using thirty second exposures, and distant enough for no parallax. A first image of the field shows Hispania 308 blended with a nearby star, within the two pixel margin of error. In the second frame the asteroid de-blends from the star and the system declares this as a new object. This triggered a follow-up response and the system took deep field images of the asteroid. This has happened with other asteroids on a few evenings, providing a continuous test of the system.

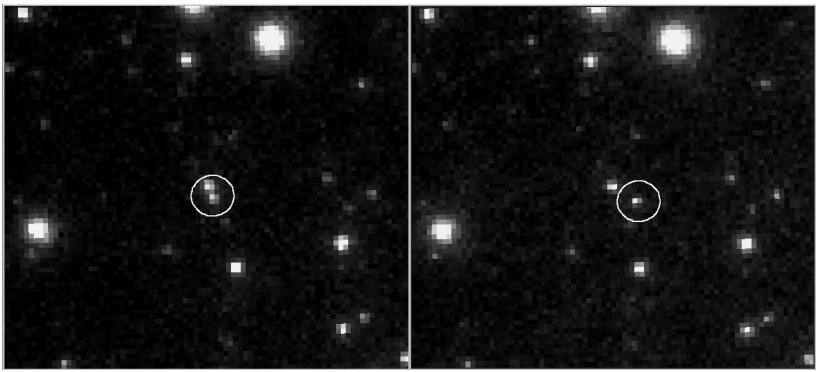

Figure 3-2 Hispania 308 flyby caught by the RAPTOR system through a closed loop alert. The left image was taken 10/17/03 04:40:6.53 UT. The right image was taken 04:41:7.13 UT on the same evening.

## 4 FUTURE DEVELOPMENT

TALON and the RAPTOR system and are still in the fledgling years and are just now beginning to yield real results. There is interest from outside institutions in tying into our system. The system was conceived and implemented with the idea that additional systems can be added seamlessly into TALON. There is some consideration being given to implementing Common Object Request Broker Architecture (CORBA), through one of the many great libraries currently available. Some of these are cross platform, well debugged, and if configured properly, address all of the distributed sensor network issues. By using CORBA we might eliminate most future errors in network code that we have not foreseen in our own programming.

However any client/server model has a few weaknesses that will need to be addressed should TALON grow large in size. The client/server model requires many trips over the network to assure one transaction, particularly if the data stream being handled is large. Each trip creates traffic and consumes bandwidth and with a great number of transactions bandwidth requirements could quickly be exceeded resulting in low system performance. Since getting on some of these target objects quickly is a requirement, this could be a serious issue. The client server model also needs the connection to be stable over the full duration of the transaction; if it goes down then there is a loss of data. Currently checks in the code verify the quality of information, but if data is lost then it has to be resent or the bad data deleted. The model requires that the capability of the network be taken into consideration during design; if the design is inaccurate then the performance of the whole system can suffer. There are solutions being implemented by engineers in the distributed sensor network field and TALON is looking into better programming options for the network code in the event we grow too big.

A section of the code currently exists that allows for a connection to be made to and from SkyDOT, an all sky database focused on temporal information. Once machine learning tools begin to mine out objects of interest from the database on regular intervals, then target of opportunity alerts will be generated and fed to clients wanting to do studies of these objects. The alert clients will store a schedule of possible targets that can be fed into the telescopes schedule system manually or automatically.

It is very expensive for one organization to attempt to garner enough resources to provide full sky coverage every night year around; however, several small programs can unite together and synchronize observations. Full sky coverage is an important issue and we are encouraging other autonomous systems to connect to the TALON system. Redundant coverage of the sky would also help eliminate lost data for sections of the sky that are obscured due to weather or instrument outages, addressing the DSN issue of mosaic coverage and fault tolerance. By combining efforts there is a better chance of mining many more extraordinary events and providing verification and follow up each night.

## 5 CONCLUSIONS

Confidence in sensor data is always better if it can be independently verified by other sensors. There will always be data drop outs, system problems and instrumental errors, but if a sensor is part of a redundant group, then the chances of the data being corrupted by errors is dramatically reduced. A network of redundant telescopes has a greater chance of producing quality results without being disadvantaged by systematic errors. When constructing autonomous robotic systems capable of identifying new objects in real-time, then independent verification and stereoscopic observation are necessary to reduce false positives and wasted follow-up observations. A communication network between the stereoscopic components becomes a key element. An intelligent network capable of tying together multiple telescopic resources to provide verification, perform data sorting, perform collaborative signal analysis, and trigger fast follow-up observations will prove to be an invaluable resource in monitoring the night sky. As more resources are tied together in a common network hub, additional confidence in observations can be achieved along with greater overall coverage of the sky without excessive expense to smaller projects. Redundant and synchronized observations by groups of telescopes can provide a stronger depth of data and fault tolerance for each night's observation. While originally slated as the means to create closed loop operation of the RAPTOR system, TALON now provides the hub necessary for tying together multiple telescopic resources and providing communication between the resources and opening a doorway to cooperative autonomous robotic observations.